\documentclass[preprint,5p,twocolumn,amssymb,amsmath,amsfonts,bbold]{elsarticle}
\usepackage{graphicx}
\usepackage{amsmath}
\usepackage{amsfonts}
\usepackage{amssymb}
\usepackage{bbm}

\begin{document}

\title{Pair production at the edge of the QED flux tube}
\author[elte,wigner_rmi]{D\'aniel Ber\'enyi}

\author[wigner_szfi]{S\'andor Varr\'o}

\author[wmu]{Vladimir V. Skokov}

\author[wigner_rmi]{P\'eter L\'evai}

\address[elte]{Lor\'and E\"{o}tv\"{o}s University, H-1117, Budapest, Hungary}
\address[wigner_rmi]{Wigner RCP, Institute for Particle and Nuclear Physics, P.O. Box 49, Budapest 1525, Hungary}
\address[wigner_szfi]{Wigner RCP, Institute for Solid State Physics and Optics, P.O. Box 49, Budapest 1525, Hungary}
\address[wmu]{Department of Physics, Western Michigan University,  1903 W. Michigan Avenue, Kalamazoo, MI 49008}

\begin{abstract}
We investigate the process of Abelian pair production in the presence of strong inhomogeneous and time-dependent external electric fields. 
The spatial dependence  of the external field is motivated by a non-Abelian color flux tube in heavy-ion collisions. We show that the inhomogeneity significantly increase the particle yield compared to that in the commonly used models with a constant and homogeneous field. 
Moreover our results indicate that in contrast to the latter, most of the particles are produced at the interface  of the field profile
 in accordance with Heisenberg's prediction.
\end{abstract}

\maketitle

\section{Introduction}

Recently, pair production from vacuum is  gaining  interest from both theorists and experimentalists. In the Abelian (QED) case it is considered to be the final frontier of high-energy laser experiments. While the attainable energy density of today experiments is still orders of magnitudes below the threshold defined 
by the critical field $E_{cr} = \frac{m^2 c^3}{e\hbar}\approx 1.32 \cdot 10^{18} \mathrm{\frac{V}{m}}$~\cite{Schwinger}, the scale above which pair production is sizable, the development of technology is remarkably sustaining its exponential growth both of energy density and of the frequency of laser pulses. The recent proposals in this area promise to improve these parameters further in the coming years \cite{TajimaMourou1, TajimaMourou2}. The high interest of research in this area is signaled by the number of high intensity laser experiments under commissioning, construction and planning, such as the Extreme Light Infrastructure (ELI). For a comprehensive list see Ref.~\cite{DiPiazzaReview}.

Another motivation comes from ultrarelativistic heavy-ion collisions performed at Relativistic Heavy Ion Collider (RHIC) and Large Hadron Collider (LHC). While the microscopic mechanism of hadron production is still not fully understood there is continued effort to develop and improve models that explain experimental data. The investigation of one of the most precisely measured observable, the transverse momentum spectra of produced hadrons, led to the development of a family of models that center around the concept of chromoelectric flux tubes ('strings'). These tubes connect the quark and diquark constituents of colliding protons~\cite{FRITIOF,HIJ,RQMD}. 
When the sources of these fluxtubes separate, the field energy increases until the threshold of pair production is reached and new quark-antiquark and diquark-antidiquark pairs are created. Such models can describe experimental data successfully at low $p_T$, $p_T < 2-3$ GeV, while at higher $p_T$ perturbative QCD-based models work well~\cite{FieldFeyn,Wang00,Yi02}. However the insight into the microscopical details is quite limited, because these models usually assume homogeneous and often static approximations of field strength \cite{StringFragmentation1, StringFragmentation2}, while in reality the collision and successive events take place on short timescales, and finite size effects may also play an important role.


During the past decades, kinetic description was formulated to describe pair production in arbitrary space-time dependent fields in the Abelian case both for fermions \cite{BirulaDHW} and bosons \cite{ScalarDHW, ScalarDHW2} and for the non-Abelian case for quarks \cite{Ochs:1998qj,QuarkWignerFunction,Skokov:2007gy} and gluons \cite{GluonSourceTerm}. While remarkable analytic progresses have been made \cite{BirulaDHWPrecession, IldertonLightFront, Fedotov}, calculating pair production in arbitrary space-time dependent fields is analytically unmanageable and also numerically very demanding. For this reason, usually only the homogeneous and often only the time independent cases are investigated and used \cite{BlaschkeApp, RuffiniApp1, RuffiniApp2}. Still, even in the most simple cases the interplay of field parameters on particle spectra and yield is so complex \cite{Skokov:2007gy,Levai:2009mn,HebenstreitSubcycle, OrthaberMultiscale} that before building full featured simulation frameworks, the independent role of field parameters should be understood.

In the case of laser experiments, the pair production is the shortest  process to take place and followed by other processes, like back-reaction, radiation reactions etc. While the laser-plasma interaction simulation packages developed today are focused on the physical reactions attainable with todays' laser energies, they expected to form the core of later full featured simulation environments, as it happened in particle and nuclear physics. In this context it is extremely important to fully understand the interplay of parameters that influence pair production observables so that future facilities can be planned for such measurements based on these simulation frameworks. Also, in heavy-ion physics pair production is one of the early stage processes and followed by multi particle interaction, energy loss, thermalization and fragmentation into hadrons that may be more or less important depending on the particle energy. The common feature in these two scenarios is that it is not clear how important the secondary processes are compared to the initial particle creation. 

In this paper, we use the three dimensional Dirac-Heisenberg-Wigner evolution equations for QED to model pair production in an inhomogeneous external electric field. The theoretical details are summarized in Section 2. The numerical method is outlined in Section 3. Our results for the longitudinal and transverse spectra are presented in Sections 4 and 5 and summarized in Section 6.

\section{The Dirac-Heisenberg-Wigner formalism}

Pair production is a quantum phenomena and thus needs a proper quantum description. Also, to account for extreme scales needed for these processes, a relativistic description is necessary. An insightful description in phase-space is available in the form of singe-time relativistic one-particle Wigner function formalism \cite{BirulaDHW}. The Wigner function is a quantum generalization of the classical one-particle distribution function. However, in contrast to the latter the Wigner function does not posses a probabilistic interpretation  in the strict sense. The Wigner function describes the quantum interference of negative energy states in the form of negative values it may posses, but these are restricted to non-connected areas with extent of the order of few $\hbar$. By integrating with a gaussian envelope function that mimics a classical measurement one ends with a classically interpretable distribution function.

While we would like to describe non-Abelian pair production, it was shown already \cite{Skokov:2007gy} that there is a strong Abelian dominance that enables us to use the U(1) formalism of QED instead of the more complicated SU(N) models. 

The definition of the one particle Wigner function in terms of the wave function is:
\begin{equation}
\begin{split}
W(\vec{x}, \vec{p}, t)=-\frac{1}{2} \int d^3 s e^{-i\vec{p}\vec{s}} & \\ \langle 0| e^{-ie \int\limits_{-1/2}^{1/2}\vec{A}(\vec{x}+\lambda\cdot\vec{s}, t)\vec{s}d\lambda} & \left[ \Psi( \vec{x} + \frac{\vec{s}}{2}, t), \bar{\Psi}( \vec{x} - \frac{\vec{s}}{2}, t) \right] |0\rangle \;,
\end{split}
\end{equation}
where the Wilson line factor is to guarantee that $p$ is the proper eigenvalue of the kinetic momentum operator. 
The evolution equations are derived in the temporal gauge in Refs. \cite{BirulaDHW, AlkoferIDHW}.
 In this approach, the external field is treated as classical. This approximation is justified by the strong field strength needed for pair production in laser experiments. In  high-energy heavy-ion collisions the gluon density is also high~\cite{McLerran:2010uc} thus it can be approximated by its classical expectation value.


The evolution equations of the Wigner function contain three non-local differential operators defined as integrals in Fourier space. By virtue of the gradient expansion, the evolution equations can be approximated by series in configuration space that include increasing gradients of external fields multiplied by increasing order of momentum derivates which act on the Wigner function. In our case, we restrict ourselves to electric fields. There is no magnetic contribution to the spatial gradients and to the momentum. We also choose a linearly polarized electric field in the $z$ direction with a transverse ($x$) spatial dependence. With this electric field the time derivate operator becomes:
\begin{equation}
D_t = \partial_t + eE_z(x, t)\cdot\partial_{p_z}\;.
\end{equation}

 This is an exact expression for the chosen electric field  because it has no component in the direction of inhomogeneity and in the direction of the electric field there is no inhomogeneity, so the higher momentum derivates vanish. This is in contrast to \cite{AlkoferIDHW, AlkoferIDHW2} where the inhomogeneity is in the direction of the electric field and the authors show that the naive gradient approximation breaks down.

In this case,  the evolution equations for the 16 real components of the Wigner function simplify to:
\begin{align}
	&D_t \mathbbm{s}				& & & &    & &-& & 2\vec{p} \cdot \mathbbm{\vec{t}}_1         &=& 0\;,& \label{EqDHWStart} \\
	&D_t \mathbbm{p}				& & & &    & &+& & 2\vec{p} \cdot \mathbbm{\vec{t}}_2         &=& 2m\mathbbm{a}_0\;,& \\
	&D_t \mathbbm{v}_0			& &+& & \partial_x \mathbbm{v}_x & & & &             &=& 0\;,& \\
	&D_t \mathbbm{a}_0			& &+& & \partial_x \mathbbm{a}_x & & & &             &=& 2m\mathbbm{p}\;,& \\
	&D_t \mathbbm{\vec{v}}		& &+& & \partial_x\mathbbm{v}_0 & &+& & 2\vec{p}\times \vec{\mathbbm{a}}  &=& -2m\vec{\mathbbm{t}}_1\;,& \\
	&D_t \mathbbm{\vec{a}}		& &+& & \partial_x\mathbbm{a}_0 & &+& & 2\vec{p}\times \vec{\mathbbm{v}}  &=& 0\;,& \\
	&D_t \mathbbm{\vec{t}}_1	& &+& & \vec{\nabla}_x \times \mathbbm{\vec{t}}_2 & &+& & 2\vec{p}\mathbbm{s}  &=& 2m\mathbbm{v}\;,& \\
	&D_t \mathbbm{\vec{t}}_2	& &-& & \vec{\nabla}_x \times \mathbbm{\vec{t}}_1 & &-& & 2\vec{p}\mathbbm{p}  &=& 0\;,& \label{EqDHWEnd}
\end{align}
where $\vec{\nabla}_x \times \mathbbm{\vec{t}}_i$ is understood as the vector $\left(0, -\partial_x \mathbbm{t}_{iz}, \partial_x \mathbbm{t}_{iy}\right)$. 
Only two derivatives remain,  namely the momentum derivatives parallel to the electric field and the spatial derivates transverse to the electric field. While it is tempting to use the method of characteristics to decouple the flow term derivates as in the homogeneous case, there is no gain in this case, because the implied momentum will be different at different coordinates and the two directions remain coupled by the spatial gradients.

The initial conditions 
\begin{eqnarray}
\mathbbm{s}(\vec{x}, \vec{p}, t=-\infty) &=& -\frac{2m}{\omega(\vec{p})}\;,\\ 
\vec{\mathbbm{v}}(\vec{x}, \vec{p}, t=-\infty) &=& -\frac{2\vec{p}}{\omega(\vec{p})}\;,
\end{eqnarray}
where $\omega^2(\vec{p})=m^2+p_x^2+p_y^2+p_z^2$, 
corresponds to the vacuum in-state. 
 The final observable that one may consider as the 'density of pairs' is the energy density, that is a linear combination of the mass density and the current density:
\begin{equation}
\epsilon = m\mathbbm{s} + \vec{p}\cdot\vec{\mathbbm{v}}\;.
\end{equation}
In the homogeneous case the DHW equations reduce to the quantum Vlasov equation with one of its components  $f$ describing the density of pairs~\cite{Levai:2009mn}:

\begin{align}
	&\partial_t f = eE(t)\frac{\sqrt{m^2+p_T^2}}{\omega(p)}\cdot v \;,\\
	&\partial_t u = 2\omega(p)\cdot v \;,\\
	&\partial_t v = eE(t)\frac{2\sqrt{m^2+p_T^2}}{\omega(p)}\cdot (1-2f) - 2\omega(p)\cdot u\;
\end{align}
with $p_T^2=p_x^2 + p_y^2$,  $p_z = p_{z_0} - eA(t)$  and the initial conditions  $f=u=v=0$ for all $\vec{p}$. Also, $f$ and $\epsilon$ are related by:
\begin{equation}
f = \frac{\epsilon}{4\omega} + \frac{1}{2}\;.
\label{feq}
\end{equation}
For the Sauter field $E(t) = E_0\cosh(t/\tau)^{-2}$ the analytic form of the asymptotic pair density is known (in fact for arbitrary spin particles) \cite{Kruglov}:

\begin{align}
&f(\vec{p}, t=\infty)=\nonumber \\
&\frac{\sinh(\pi (\theta - \mu_+ + \mu_-)) \sinh(\pi (\theta + \mu_+ - \mu_-))}{\sinh(2\pi\mu_+) \sinh(2\pi\mu_-)}\;,
\end{align}
where
\begin{eqnarray}
\mu_{\pm} &=& \frac{1}{2} \tau \sqrt{ (p_{z0} \pm eE_0\tau)^2 + p_T^2 + m^2 }\;,\\
\theta &=& eE_0\tau^2\;.
\end{eqnarray}
We use this result to validate our numerics and to assess the effect of inhomogeneity.  We chose the following 'plateau' field, that models the cross-section of a chromoelectric flux tube in the Abelian limit, assuming homogeneous field in the middle and exponential decay at the edges:
\begin{align}
&E(x, t)=E_0\cosh(t/\tau)^{-2}\\
&\times\frac{1}{2}\left( 1 - \tanh\left( \frac{x+R}{r}\right)\tanh\left(\frac{x-R}{ r }\right)\right)\;. 
\end{align}
 Thus we have two parameters for the spatial dependence: $R$ the width of the plateau (or radius of the flux tube) and $r$ describing the steepness of the gradient at the edge. In the time direction we use the Sauter field so we can expect to reproduce the homogeneous result at $x\approx0$ when $R \gg r$. We restrict ourselves to a one-dimensional inhomogeneity in three dimensions. Practically this field is an infinite homogeneous flux plane possessing a finite extent and gradient only in the direction perpendicular to this plane.
 
\section{Numerical method}

We solve the equation system (\ref{EqDHWStart}-\ref{EqDHWEnd}) by explicit finite difference integration in the time direction with a usual 8th order Runge-Kutta stepper and account for the spatial derivates with pseudo-spectral collocation over the rational Chebyshev polynomial basis \cite{BoydRatCheb}. These polynomials resolve doubly infinite ranges with a user defined scale. The reason for the choice of this method is that the Wigner function is free of non-analyticities and for this class of functions the spectral method has superior convergence rates over finite difference techniques. Another advantage is that the collocation points  coincide with those optimal for the integration quadrature . Thus the integrations can be carried out with no further efforts  in spectral space with a Clenshaw-Curtis quadrature modified for the rational Chebyshev basis set \cite{BoydRatCheb2}.

Within the pseudo-spectral method, our differential operators turn into dense matrices. Because the operators acting on the Wigner-function components are time independent, we can solve the equations in spectral space, the back and forward transformation in each time step is not necessary. We use this method also because it can be easily extended and programmed on graphical processors (GPUs).

The free scale parameters of the rational Chebyshev polynomials should be estimated before calculations. A good choice for the $x$ direction is in the order of $2R+r$, while for the scale of the longitudinal momentum $p_z$ the integral of the electric field sets the scale: $\int\limits_{-\infty}^{\infty}eE(t)dt = eA(-\infty)-eA(\infty)$. 

 \begin{figure}
 \centerline{%
 \rotatebox{-90}{\includegraphics[height=8.0truecm]{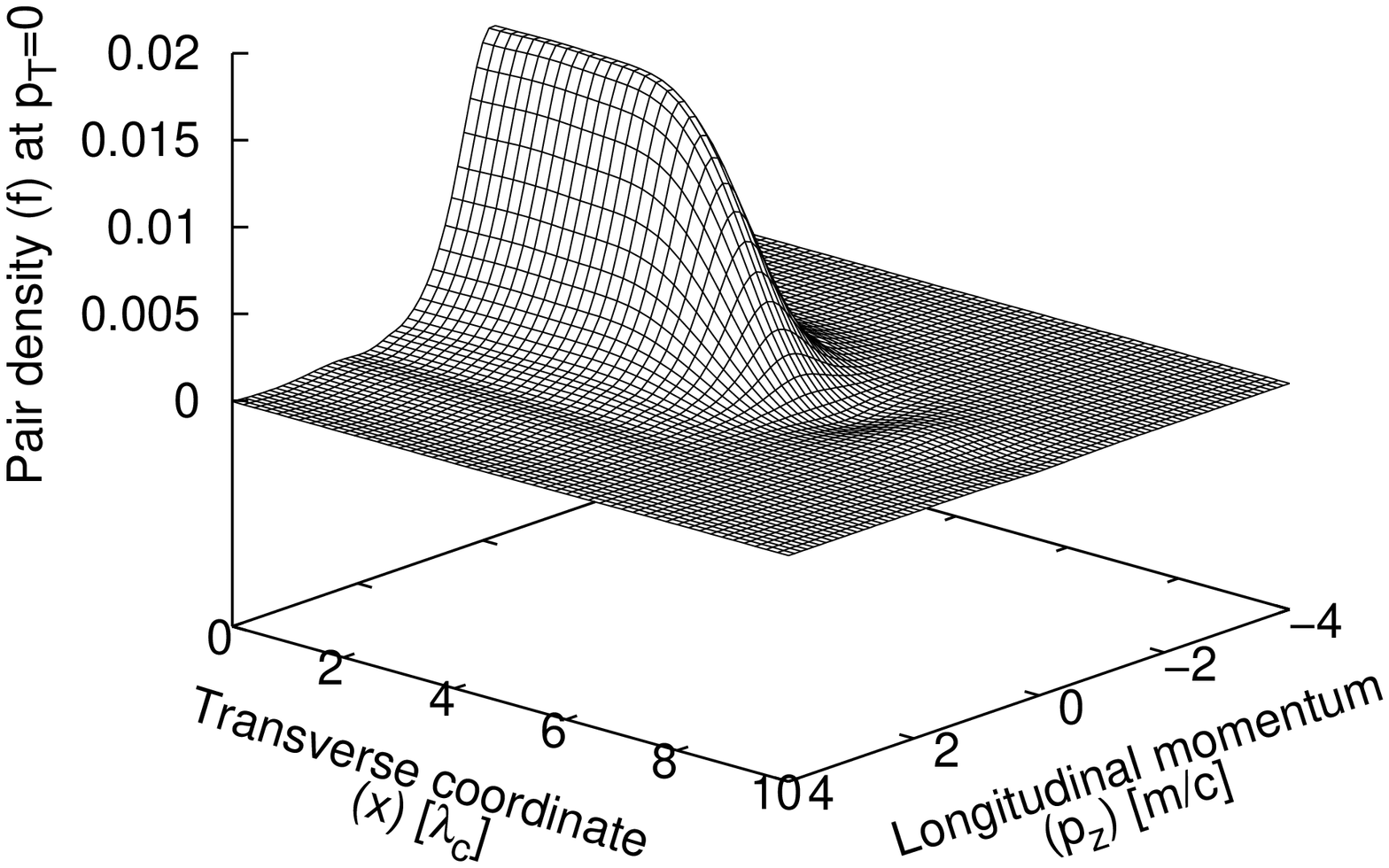}}}
 \caption{Phase space view of the asymptotic pair density with parameters: $E_0 = 0.5E_{cr}, R=5\lambda_c, r=\lambda_c, \tau = 0.3\lambda_c/c$. Note that the $E$ field has the steepest gradient at $z=5\lambda_c$.}
 \label{3D1}

 \centerline{%
 \rotatebox{-90}{\includegraphics[height=8.0truecm]{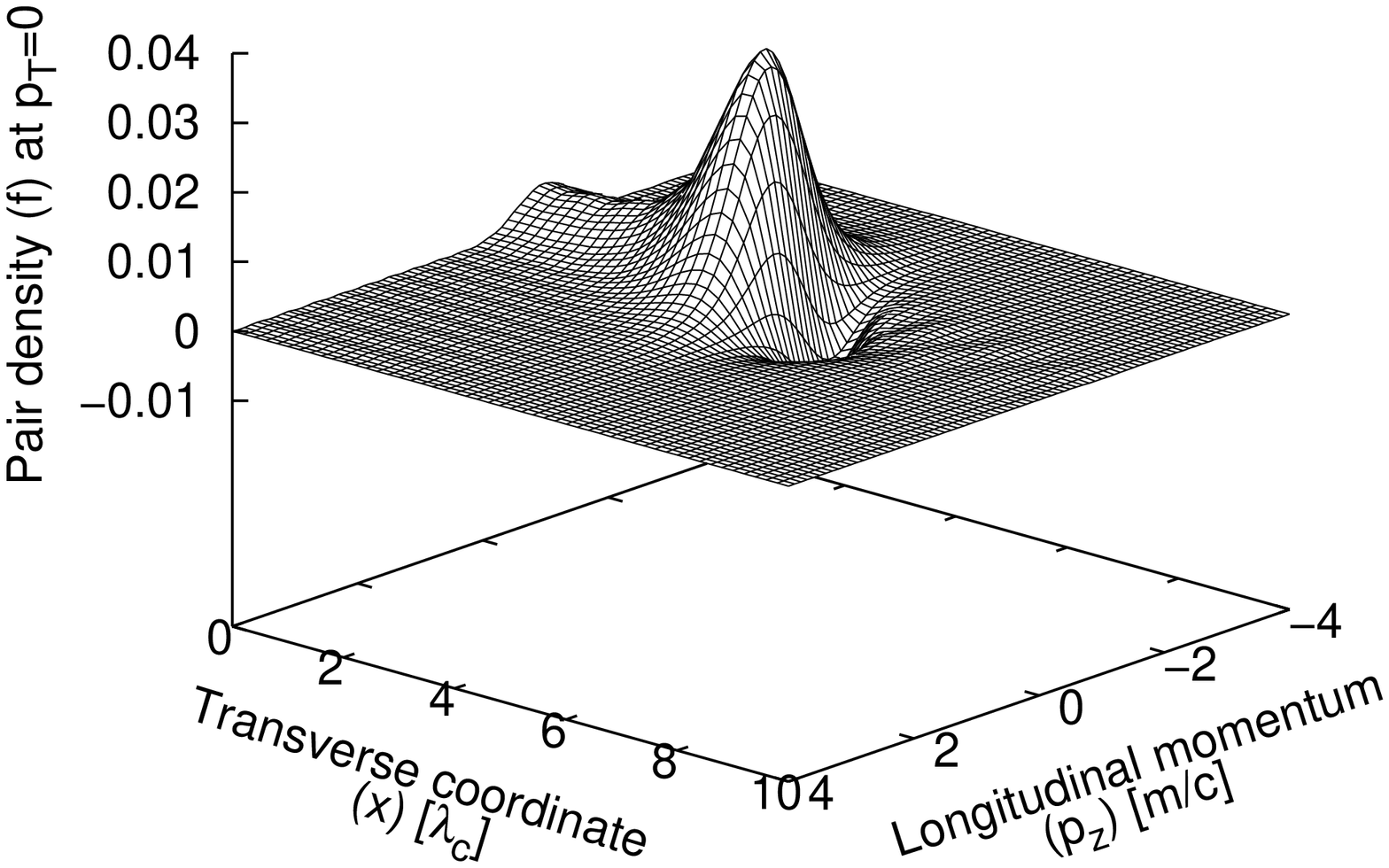}}}
 \caption{Phase space view of the asymptotic pair density with parameters: $E_0 = 0.5E_{cr}, R=5\lambda_c, r=\lambda_c, \tau = 2\lambda_c/c$. Note that the $E$ field has the steepest gradient at $x=5\lambda_c$.}
 \label{3D2}
 \end{figure}
 
 \begin{figure}
 \centerline{%
 \rotatebox{-90}{\includegraphics[height=8.0truecm]{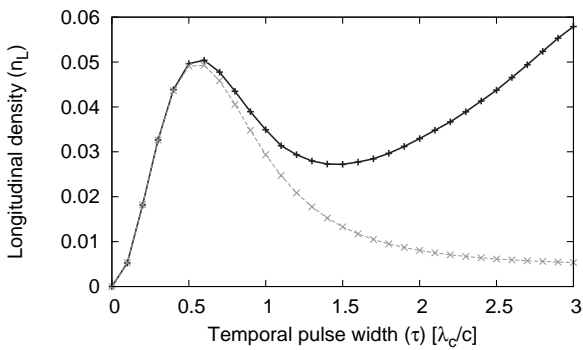}}}
 \caption{Pulse width ($\tau$) dependence of particle yield. Solid black line: inhomogeneous model, dashed grey line: homogeneous reference.}
 \label{taudep}
 \end{figure}
 
 \begin{figure}
 \centerline{%
 \rotatebox{-90}{\includegraphics[height=8.0truecm]{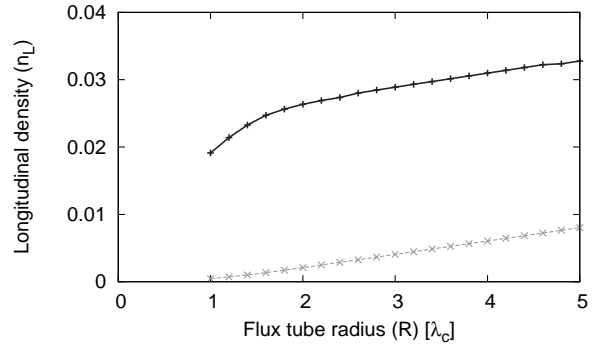}}}
 \caption{Flux tube radius ($R$) dependence of the particle yield. Solid black line: inhomogeneous model, dashed grey line: homogeneous reference. Note the same slope.}
 \label{Ldep}
 \end{figure}

\section{Longitudinal spectra}
At first we restrict ourselves to zero transverse momentum. It is known that the $p_T$ only acts as an additional mass term regarding the particle yield; increasing it results in an exponential decay of the pair production.

The 3-d plots on Figs. \ref{3D1}. and \ref{3D2}. we show the asymptotic ($t\to \infty$) pair density $f$, as defined by Eq. (\ref{feq}), as a function of  the transverse coordinate $x$ and longitudinal ($z$) momentum for two different values $\tau=0.3\lambda_c/c$ and $\tau=2\lambda_c/c$. We note that the first value corresponds to the local maxima of the homogeneous Sauter model in $\tau$ space. 
The middle of the plateu behaves like the homogeneous solution, decreasing with increasing $\tau$, but the inhomogeneous edge tends to increase the pair density in its vicinity proportional to the pulse width. Next to the edge a negative region develops that signifies the admixture of antiparticles. The less steeper the gradient, the less negative density appears. We note that $f$ is not a component of the Wigner function, therefore there are no bounds on the values it may take.

To have a scalar quantity to compare to the homogeneous case we calculate the particle yield integrated in the transverse coordinate and longitudinal momentum:
\begin{equation}
n_L = \int\limits_{-\infty}^{\infty} \mathrm{d}x \frac{\mathrm{d}p_z}{2\pi} f(x, p_x=0, p_y=0, p_z)\;. 
\end{equation}
This quantity can be compared to the homogeneous case by calculating it with the same $E(x, t)$ field as in the inhomogeneous one and setting $p_T=0$ and performing the same integrals in $x$ and $p_z$. First we plot $n_L$ as a function of the Sauter pulse width $\tau$, while keeping $R$ and $r$ fixed on Fig. \ref{taudep}. We see that the homogeneous and inhomogeneous results for small temporal widths coincide as expected as there is no time for the particles to be created by the inhomogeneity. The two curves reach a local maximum as in the homogeneous case, but they start to separate. For large $\tau$ both curves are expected to be proportional to it, and finally will be approximated with the form of $\propto \tau \times n_{static}$, where $n_{static}$ is the constant static solution (for the homogeneous field it is proportional to the Schwinger formula). We find, that the onset of this approximation happens earlier for the inhomogeneous configuration and thus the observed particle yield is more than a magnitude larger than in the homogeneous case.


If we fix pulse width $\tau=2\lambda_c/c$ and gradient $r=\lambda_c$ and vary $R$, we again expect a linear proportionality since we are changing the interaction volume. We indeed find a linear relation (Fig. \ref{Ldep}) and also the slopes turn out to be the same within 5\%. This means that the homogeneous and inhomogeneous solutions are almost identical in the sense of volumetric scaling. This together with Figs. \ref{3D1}-\ref{3D2} implies that inhomogeneous pair production is a surface effect as it was already predicted by Heisenberg in 1934 when he calculated the fluctuation of a charge in QED and found that it was $\propto V^{\frac{2}{3}}$ or the surface of the interaction volume \cite{HeisenbergPrediction}. This  is in a disagreement to what one may conclude from the constant-homogeneous string models usually applied in heavy ion physics.

\section{Transverse spectra}

\begin{figure}
\centerline{%
\rotatebox{-90}{\includegraphics[height=8.0truecm]{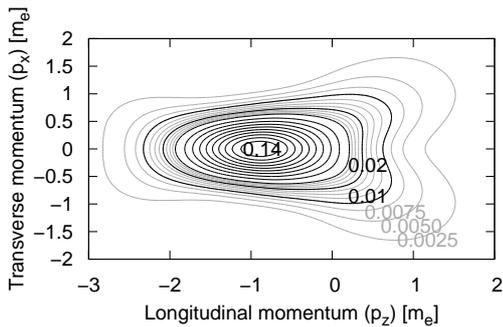}}}
\caption{Transverse and longitudinal spectra of pair density with parameters as in Fig. \ref{3D2}. Note that the right side of the distribution is wider than the left. The distribution is accelerated to the left.}
\label{TLspectra}
\end{figure}

The $p_y$ momentum is fully conserved in this setup so it is just an additional mass term and therefore set to 0 in the following. We integrate the $x$ coordinate and plot the $p_z-p_x$ momentum spectra in Fig. \ref{TLspectra}. In the homogeneous case it would be a simple radially symmetric peak as larger momenta are exponentially suppressed. However in this case particles are also created by the inhomogeneity and they depart from the main peak and this increases the production rate in the transverse direction. The earlier the particles produced the further they get from the center and with the additional effect of accelerating in the $E$ field this gives a triangular shape to the distribution.

\begin{figure}
\centerline{%
\rotatebox{-90}{\includegraphics[height=8.0truecm]{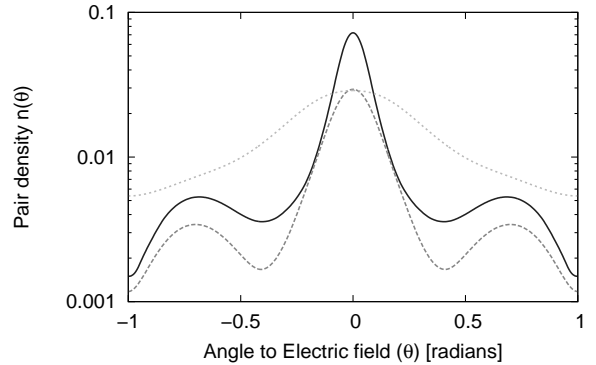}}}
\caption{Angular dependence of the created particles calculated from Fig. \ref{TLspectra}. $\theta$ is zero in the direction of the electric field. Solid black: $\tau=3 \lambda_c/c$, dashed dark grey: $\tau=2 \lambda_c/c$, dotted light grey: $\tau=1 \lambda_c/c$.}
\label{Tangulard}
\end{figure}

The angular distribution of pairs is an important observable. It can be computed by the momentum integral in polar coordinates  
\begin{equation}
n(\theta) = \int\limits_{0}^{\infty} \frac{\mathrm{d}p \mathrm{d}x}{2\pi} p f(x, p_x = p\sin(\theta), p_y = 0, p_z = p\cos(\theta) ).
\end{equation}
The resulting distribution can be seen in Fig. \ref{Tangulard}. For the homogeneous case this would be a simple peak as illustrated by the case $\tau=\lambda_c/c$, but the excess particles provided by the inhomogeneity for $\tau=2\lambda_c/c$ and $\tau=3\lambda_c/c$ give rise to two side peaks while the central peak shrinks due to its increasing distance from the origin. The appearance of side peaks is remarkably similar to the predicted bifurcation for squeezed states in quantum optics, see Ref.~\cite{VS_bifurcation}.

\section{Summary}

We investigated the Abelian pair production in inhomogeneous external electric field with a shape motivated by the color flux tubes in heavy ion collisions. We found that the number of particles created can be significantly underestimated by the homogeneous models. Moreover we showed evidence that the results obtained with homogeneous string models can be conceptually misleading and a proper description including finite size effects may needed. Such a description may provide microscopical model for the empirical parameters. We also presented the widening of the transverse spectra as a potential discriminant of homogeneous and inhomogeneous processes.

\section*{Acknowledgments}
D. B. and P. L. are grateful for R. Alkofer and C. Kohlf\"urst for discussions.
This work was supported in part by the Hungarian OTKA Grants No. 77816, No. 104260, No. 106119.


\begin{thebibliography}{1}
\bibitem{Schwinger} 
  J.~S.~Schwinger,
  Phys.\ Rev.\  {\bf 82}, 664 (1951).

\bibitem{TajimaMourou1}
T.Tajima, G. Mourou,
	Phys. Rev. ST Accel. Beams Vol. 5, (2002) 031301.
	
\bibitem{TajimaMourou2}	
G. Mourou, T. Tajima,
	Science 331, (2011) 41-42.
	
\bibitem{DiPiazzaReview}
A. Di Piazza et. al,
Rev. Mod. Phys. vol. 84, (2012) 1177.

\bibitem{FRITIOF} 
  B. Andersson {\it et al.}, 
    Phys. Rep. {\bf 97} (1983) 31;
    Nucl. Phys. {\bf B281} (1987) 289; 
    Z. Phys. {\bf C57} (1993) 485.

\bibitem{HIJ}
  X.N. Wang, M. Gyulassy, 
    Phys. Rev. {\bf D44} (1991) 3501;
    Comput. Phys. Commun. {\bf 83} (1994) 307.

\bibitem{RQMD} 
  H. Sorge,
  Phys. Rev. {\bf C52} (1995) 3291.

\bibitem{FieldFeyn}
  R.D. Field, {\it Application of Perturbative QCD},
  Addison-Wesley, 1989.

\bibitem{Wang00}
  X.N. Wang, 
    Phys. Rev. {\bf C61} (2000) 064910.

\bibitem{Yi02}
  Y. Zhang {\it et al.}, Phys. Rev. {\bf C65} (2002) 034903.

\bibitem{StringFragmentation1}
V. Topor Pop et. al.,
	Phys. Rev. {\bf C86} (2012) 044902.

\bibitem{StringFragmentation2}
V. Topor Pop et. al.,	
	[arXiv:1306.0885 [hep-ph]].

\bibitem{BirulaDHW}
I. Bialynicki-Birula, P. Gornicki, J. Rafelski,
	Phys. Rev. {\bf D44 } (1991) 1825-1835.
	
\bibitem{ScalarDHW}
C. Best, P. Gornicki, W. Greiner,
	Annals Phys. 225 (1993) 169-190.
	
\bibitem{ScalarDHW2}
S. Varr\'o, J. Javanainen,
	J. Opt. B: Quantum Semiclass. Opt. {\bf 5} (2003) S402-S406.

\bibitem{Ochs:1998qj} 
  S.~Ochs and U.~W.~Heinz,
  Annals Phys.\  {\bf 266}, 351 (1998)
  [hep-th/9806118].
	
\bibitem{QuarkWignerFunction}
A.V. Prozorkevich, S.A. Smolyansky, S.V. Ilyin,
"Progress in Nonequilibrium Green's Functions II", Eds. M. Bonitz and D. Semkat, World Scientific (2003), p. 401.

\bibitem{Skokov:2007gy} 
  V.~V.~Skokov and P.~Levai,
  Phys.\ Rev.\ D {\bf 78}, 054004 (2008)
  [arXiv:0710.0229 [hep-ph]].

\bibitem{GluonSourceTerm}
D. D. Dietrich, G. C. Nayak, W. Greiner,
	Phys. Rev. {\bf D64} (2001) 074006.


	
\bibitem{IldertonLightFront}
F. Hebenstreit, A. Ilderton, M. Marklund,
	Phys. Rev. {\bf D84} (2011) 125022.
	
\bibitem{BirulaDHWPrecession}
I. Bialynicki-Birula, Lukasz Rudnick,
	Phys. Rev. {\bf D83} (2011) 065020.

\bibitem{Fedotov}
	A.M. Fedotov, A.A. Mironov,
	arXiv:1310.7258
	
\bibitem{BlaschkeApp}
D. B. Blaschke et al.,
	Eur. Phys. J. {\bf D55} (2009) 341-358.
	
\bibitem{RuffiniApp1}
R. Ruffini, S. Xue,
	Phys. Lett. {\bf B696} (2011) 416-412.

\bibitem{RuffiniApp2}
A. Benedetti, R. Ruffini, G. Vereshchagin,
	Phys. Lett. {\bf A377} (2013) 206-215.
	
\bibitem{Levai:2009mn} 
  P.~Levai and V.~Skokov,
  Phys.\ Rev.\ D {\bf 82}, 074014 (2010)
  [arXiv:0909.2323 [hep-ph]].
	
\bibitem{HebenstreitSubcycle}
F. Hebenstreit, et. al.,
	Phys. Rev. Lett. 102 (2009) 150404.
	
\bibitem{OrthaberMultiscale}
M. Orthaber, F. Hebenstreit, R. Alkofer,
	Phys. Lett. {\bf B698} (2011) 80-85.
	
\bibitem{AlkoferIDHW}
F. Hebenstreit, R. Alkofer, H. Gies,
 	Phys. Rev. {\bf D82 } (2010) 105026.

\bibitem{McLerran:2010uc} 
  L.~McLerran,
  Prog.\ Theor.\ Phys.\ Suppl.\  {\bf 187}, 17 (2011)
  [arXiv:1011.3204 [hep-ph]].

\bibitem{AlkoferIDHW2}
F. Hebenstreit, R. Alkofer, H. Gies,
	Phys. Rev. Lett. {\bf 107} (2011) 180403.
	
\bibitem{Kruglov}
S. I. Kruglov,
	Radiat. Phys. Chem. {\bf 75} (2006) 723-728.
	
\bibitem{BoydRatCheb}
J. P. Boyd,
	J. Comp. Phys. {\bf 69} (1987) 112-142.
	
\bibitem{BoydRatCheb2}
J. P. Boyd,
	J. Sci. Comp. Vol. 2, No. 2 (1987).
	
\bibitem{HeisenbergPrediction}
W. Heisenberg,
	Sachsiche Akademie der Wissenschaften, Vol. 86, p. 317 (1934).
	
\bibitem{VS_bifurcation}
W. Schleich, R. J. Horowicz, S. Varr\'o,
	Phys. Rev. {\bf A40} 12 (1989) 7405-7408.
	





\end{thebibliography}
\end{document}